\documentclass[aps,prd,superscriptaddress,showpacs,tighten,nofootinbib]{revtex4}
\usepackage[a4paper, total={6.0in, 9in}]{geometry}
\usepackage{amsfonts,amssymb,latexsym,amsmath} 
\usepackage{bm}
\usepackage{hhline}
\usepackage{longtable}
\usepackage{array} 
\usepackage[pdftex]{graphicx}

\newcommand{\partiallr}{\overleftrightarrow{\partial}}
\newcommand{\al}{&\!\!\!\!}

\newcommand{\Zc}{Z_c(3900)}
\newcommand{\Zcp}{Z_c(4040/4025)}
\newcommand{\Zb}{Z_b(10610)}
\newcommand{\Zbp}{Z_b(10650)}
\newcommand{\gev}{\mathrm{GeV}}
\newcommand{\mev}{\mathrm{MeV}}

\newcommand{\nb}{\mathrm{nb}}

\newcommand{\be}{\begin{equation}}
 \newcommand{\ee}{\end{equation}}
\newcommand{\ba}{\begin{array}{c}}
 \newcommand{\ea}{\end{array}}
\newcommand{\bea}{\begin{eqnarray}}
  \newcommand{\eea}{\end{eqnarray}}

\begin{document}

\title{$P$-wave coupled channel effects in electron-positron annihilation}

\author{Meng-Lin Du}
\email{du@hiskp.uni-bonn.de}
\affiliation{Helmholtz-Institut f\"ur Strahlen- und Kernphysik and Bethe
   Center for Theoretical Physics, \\Universit\"at Bonn,  D-53115 Bonn, Germany}
\author{Ulf-G.~Mei{\ss}ner}
\email{meissner@hiskp.uni-bonn.de}
\affiliation{Helmholtz-Institut f\"ur Strahlen- und Kernphysik and Bethe
   Center for Theoretical Physics, \\Universit\"at Bonn,  D-53115 Bonn, Germany}
 \affiliation{Institut f\"{u}r Kernphysik, Institute for Advanced
Simulation, and J\"ulich Center for Hadron
Physics,\\D-52425 J\"{u}lich, Germany}
\author{Qian Wang}
\email{wangqian@hiskp.uni-bonn.de}
\affiliation{Helmholtz-Institut f\"ur Strahlen- und Kernphysik and Bethe
   Center for Theoretical Physics, \\Universit\"at Bonn,  D-53115 Bonn, Germany}

\begin{abstract}
$P$-wave coupled channel effects arising from the
$D\bar{D}$, $D\bar{D}^*+c.c.$, and $D^*\bar{D}^*$ thresholds in
$e^+e^-$ annihilations are systematically studied.  We provide an 
exploratory study by solving the Lippmann-Schwinger equation with 
short-ranged contact potentials obtained in the heavy quark limit.
These contact potentials can be extracted from the $P$-wave
interactions in the $e^+e^-$ annihilations, and then be employed to 
investigate possible isosinglet $P$-wave hadronic molecules. In
particular, such an investigation may provide information about 
exotic candidates with quantum numbers $J^{PC}=1^{-+}$. In the mass region
of the $D\bar{D}$, $D\bar{D}^*+c.c.$, and $D^*\bar{D}^*$ thresholds,
there are two quark model bare states, i.e. the $\psi(3770)$ and
$\psi(4040)$, which are assigned as $(1^3D_1)$ and $(3^1S_1)$ states,
respectively. By an overall fit of  the cross sections of $e^+e^-\to
D\bar{D}$, $D\bar{D}^*+c.c.$, $D^*\bar{D}^*$, we determine the
physical coupling constants to each channel and extract the pole
positions of the $\psi(3770)$ and $\psi(4040)$. The deviation of the
ratios from that in the heavy quark spin symmetry (HQSS) limit reflects 
the HQSS breaking effect due to the mass splitting
between the $D$ and the $D^*$. Besides the two poles, we also find a pole
a few MeV above the $D\bar{D}^*+c.c.$ threshold which can be
related to the so-called $G(3900)$ observed earlier by BABAR and Belle. 
This scenario can be further scrutinized by measuring the angular
distribution in the $D^*\bar{D}^*$ channel with high luminosity
experiments.
\end{abstract}

\maketitle


\section{Introduction}
Since the observation of the $X(3872)$ in 2003, numerous exotic
candidates which do not fit into the conventional quark model spectrum have been
observed in experiments.  Most of these exotic candidates appear near
some open-flavor thresholds which calls for a systematical study of the
threshold  or coupled channel effects in the relevant
channels. For the $S$-wave interaction, the most famous one is
$D\bar{D}^*$\footnote{Here and in what follows, $D\bar{D}^*$ means
$D\bar{D}^*+c.c.$ which  includes its charged conjugate partner
to form a $C$-parity eigenstate.} interaction in the isospin singlet
channel which is crucial for understanding the nature of the
$X(3872)$ \cite{Voloshin:1976ap,Tornqvist:2004qy, Tornqvist:1991ks,
Swanson:2003tb,Braaten:2007ct,Fleming:2007rp,Wang:2013kva,Baru:2015nea,Baru:2011rs}.
The corresponding coupled channel effects in the isovector channel
have also been studied to probe the structure of the $\Zc$ and the
$\Zcp$
\cite{Chen:2014afa,He:2014nya,Coito:2016ads,Zhao:2015mga,Chen:2015jwa,Prelovsek:2014swa,Aceti:2014uea,Aceti:2014kja,Guo:2013sya}
as well as their heavy flavor partners in the bottom sector, i.e.
$\Zb$ and $\Zbp$
\cite{Dias:2014pva,Sun:2011uh,Cleven:2011gp,Hanhart:2015cua,Guo:2016bjq},
which are close to the $B\bar{B}^*$ and $B^*\bar{B}^*$ thresholds,
respectively. Since the $S$-wave interaction among hadrons can form
a bound state more easily than other  partial waves, the
nearby $S$-wave threshold and the $S$-wave interaction which might
form a hadronic molecule have attracted a lot of attention. Although
an interaction in higher partial waves cannot easily  form a
bound state, it could also have moderate effects on certain observables
within the relevant energy region, especially the next
alternative option of a $P$-wave. 
 Besides the potential
$S$-wave hadronic molecules mentioned above, there are also some
vector exotic candidates which appear in $e^+e^-$ annihilation,
such as $G(3900)$ \cite{Aubert:2006mi,Pakhlova:2008zza},
$Y(4008)$\cite{Yuan:2007sj}, $Y(4260)$ \cite{Aubert:2005rm},
$Y(4360)$\cite{Aubert:2007zz}, $Y(4630)$\cite{Pakhlova:2008vn},
$Y(4660)$\cite{Wang:2007ea,Liu:2008hja} and so on.

With the availability of high-luminosity data from Belle and BESIII
for $e^+e^-$ annihilation,
it is timely to study the $P$-wave interactions between a pair of
 $S$-wave heavy-light mesons, such as $D\bar{D}$, $D\bar{D}^*$, and
$D^*\bar{D}^*$, as well as the $S$-wave interactions 
between one $S$-wave heavy-light meson and one $P$-wave heavy-light meson, such as
$D_1\bar{D}+c.c.$, $D_1\bar{D}^*+c.c.$, and $D_2\bar{D}^*+c.c.$
pairs. This will help us understand both the conventional heavy
quarkonium and the vector exotic candidates. Since in the heavy
quark limit the interaction between the two spin multiplets share
the same low-energy parameters, the study will also shed light on
the existence of other possible exotic candidates but with different
quantum numbers. However, up to now, most of the studies on the
$P$-wave or $S$-wave threshold effect in $e^+e^-$ colliders are
mainly based on the one-loop calculation
\cite{Bondar:2016pox,Liu:2014spa,Cleven:2013mka,Qin:2016spb,Dubynskiy:2006cj,Zhang:2010zv}
within some power counting schemes \cite{Guo:2010ak} or the effective
Lagrangian approach. There are also some studies
\cite{Chen:2012qq,Liu:2010xh,Achasov:2012ss,Cao:2014qna} which focus
on the one-channel case. As a result, a systematic study of
the $P$-wave as well as the $S$-wave interaction in $e^+e^-$
annihilation is called for. 

In this paper, we study the $P$-wave $D\bar{D}$, $D\bar{D}^*$, and
$D^*\bar{D}^*$ (Fig.~\ref{fig:FeynmanDiagram}) coupled channel
effects in $e^+e^-$ annihilation within the energy region $[3.70,
4.25]$~GeV. Since the probability for the creation of a pair of strange quarks is much
smaller than that of an up and$/$or down quark pair, we neglect the
strange charmed thresholds.\footnote{A further argument in 
favor of this assumption is the fact that the cross section of $e^+e^-\to D_s^{(*)}
\bar{D}_s^{(*)}$ \cite{Pakhlova:2010ek} is about one order of
magnitude smaller than those of $e^+e^-\to D\bar{D}$, $D\bar{D}^*$,
and $D^*\bar{D}^*$. } The next open flavor channel should be included, 
such as $D_1\bar{D}$ (about $4.29$~GeV),  as long as the energy exceeds the production threshold.
This is the reason why we only consider the energy below 4.25~GeV.  
As we are performing an  exploratory study, we include the short-ranged contact potential in
the heavy quark limit in addition to the conventional charmonia
without including the one-pion exchanged potential.\footnote{The
discussion of the one-pion exchange potential and the relevant
three-body channels, such as $D\bar{D}\pi$, $D\bar{D}^*\pi$ will be
included in the forthcoming work as well as the next $S$-wave
thresholds, i.e. $D_1\bar{D}+c.c.$, $D_1\bar{D}^*+c.c.$, and
$D_2\bar{D}^*+c.c.$.} In Sec.~\ref{sec:formula}, we present the
decomposition of $P$-wave heavy-light meson pair in the heavy quark
limit and the corresponding Lippmann-Schwinger equation.
Section~\ref{sec:result} contains the results and discussion.
We end with a summary in Sec.~\ref{sec:summary}.

\section{Formalism}

In this section, we present the decomposition formula of the
$P$-wave $D\bar{D}$, $D\bar{D}^{*}$, and $D^{*}\bar{D}^{*}$ with
quantum number $J^{PC}=1^{--}$ in terms of the heavy and
light degrees of freedom which are conserved,
respectively, in the heavy quark limit. Accordingly, the short-ranged
contact potentials can be obtained from this decomposition. In what
follows, the Lippmann-Schwinger equation in the calculation is
presented explicitly. \label{sec:formula}

\subsection{Decomposition of {\boldmath$P$}-wave {\boldmath$D\bar{D}$}, 
{\boldmath$D\bar{D}^{*}$}, and {\boldmath$D^{*}\bar{D}^{*}$} pairs}\label{sec:decomposition}

\begin{figure}[tb]
\begin{center}
  \includegraphics[width=0.6\textwidth]{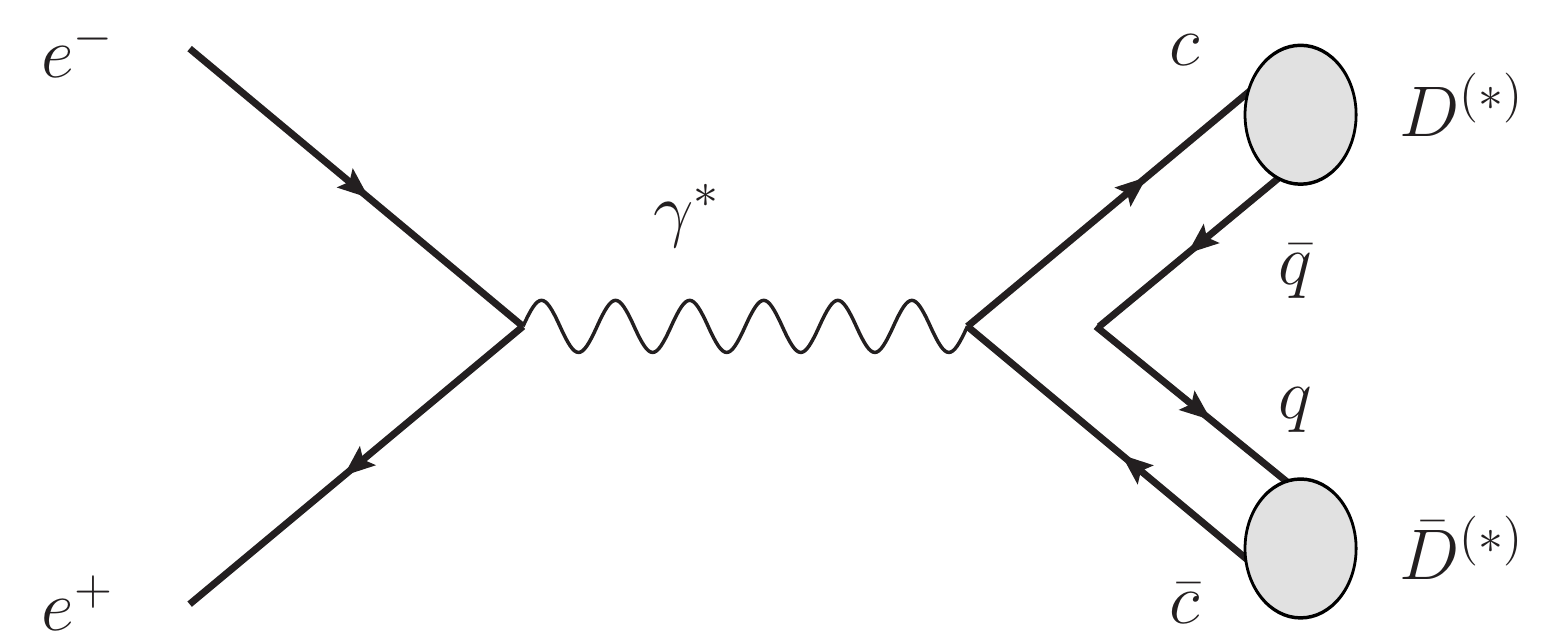}
\caption{Feynman diagram for $e^+e^-$ to a pair of
charmed mesons. } \label{fig:FeynmanDiagram}
\end{center}
\end{figure}

The decomposition of a pair of charmed mesons with a relative
orbital angular momentum $l$ reads \footnote{Here $\hat{j}=\sqrt{2 j+1}$.}
\begin{eqnarray}
\label{eq:decomposition}
|l([s_{l_{1}}s_{Q_{1}}]_{j_{1}}[s_{l_{2}}s_{Q_{2}}]_{j_{2}})_{s}\rangle_{J}
& = &
\sum_{s_{l},s_{Q},s_{q}}(-1)^{l+s_{q}+s_{Q}+J}\hat{s_{q}}\hat{s_{Q}}\hat{j_{1}}\hat{j_{2}}\hat{s}\hat{s_{l}}\left\{
\begin{array}{ccc}
s_{l_{1}} & s_{Q_{1}} & j_{1}\\
s_{l_{2}} & s_{Q_{2}} & j_{2}\\
s_{q} & s_{Q} & s
\end{array}\right\}\nonumber\\
&&\times \left\{ \begin{array}{ccc}
l & s_{q} & s_{l}\\
s_{Q} & J & s
\end{array}\right\} |(l[s_{l_{1}}s_{l_{2}}]_{s_{q}})_{s_{l}}[s_{Q_{1}}s_{Q_{2}}]_{s_{Q}}\rangle_{J},
\end{eqnarray}
where
$|l([s_{l_{1}}s_{Q_{1}}]_{j_{1}}[s_{l_{2}}s_{Q_{2}}]_{j_{2}})_{s}\rangle_{J}$ 
is the hadronic basis with $s_{Q_i}$ and $s_{l_i}$ the heavy quark
spin and the spin plus relative orbital angular momentum of the light 
degrees of freedom in the $i$th hadron,
respectively, $j_i$ is the spin of the $i$th hadron and $s$ is the
sum of them. Further, $l$ and $J$ are the relative orbital angular momentum
and total angular momentum of the two-hadron system, respectively.
 The hadronic basis can be reexpressed as Eq.~(\ref{eq:decomposition}) in terms of the heavy
and light degrees of freedom basis
$|(l[s_{l_{1}}s_{l_{2}}]_{s_{q}})_{s_{l}}[s_{Q_{1}}s_{Q_{2}}]_{s_{Q}}\rangle_{J}$,
with $s_Q$ the total spin of the heavy quark pair, $s_q$ is the total spin of the
light degrees of freedom, and $s_l$ their total spin plus relative orbital
angular momenum, $s_l = s_q+l$. Since $s_Q$ and $s_l$ are conserved,
respectively, in the heavy quark spin symmetry (HQSS) limit,
$|(l[s_{l_{1}}s_{l_{2}}]_{s_{q}})_{s_{l}}[s_{Q_{1}}s_{Q_{2}}]_{s_{Q}}\rangle_{J}$
can be simplified as $|s_{Q}\otimes s_{l}\rangle_{J}$. Using
Eq.~(\ref{eq:decomposition}), one can obtain the decompositions of
the $P$-wave charmed meson pair with $J^{PC}=1^{--}$ as
\cite{Voloshin:2012dk}
\begin{eqnarray}
|D\bar{D}\rangle_{1^{--}} & = & \frac{1}{2}|0\otimes1\rangle+\frac{1}{2\sqrt{3}}
|1\otimes0\rangle-\frac{1}{2}|1\otimes1\rangle+\frac{1}{2}\sqrt{\frac{5}{3}}|1\otimes2\rangle~,
\label{eq:1}\\
|D\bar{D}^{*}+c.c.\rangle_{1^{--}} & = & -\frac{1}{\sqrt{3}}|1\otimes0\rangle+\frac{1}{2}
|1\otimes1\rangle+\frac{1}{2}\sqrt{\frac{5}{3}}|1\otimes2\rangle~, \label{eq:2}\\
|D^{*}\bar{D}^{*}\rangle^{s=0}_{1^{--}} & = & \frac{1}{2}\sqrt{3}|0\otimes1\rangle-\frac{1}{6}
|1\otimes0\rangle+\frac{1}{2\sqrt{3}}|1\otimes1\rangle-\frac{\sqrt{5}}{6}|1\otimes2\rangle~, 
\label{eq:3}\\
|D^{*}\bar{D}^{*}\rangle^{s=2}_{1^{--}} & = &
\frac{\sqrt{5}}{3}|1\otimes0\rangle+\frac{1}{2}\sqrt{\frac{5}{3}}|1\otimes1\rangle
+\frac{1}{6}|1\otimes2\rangle~.
\label{eq:4}
\end{eqnarray}
These wave functions are normalized to one and orthogonal to each other. 
The coefficients can be written as a compact matrix
\begin{eqnarray}
g^{1^{--}} & = & \left(\begin{array}{cccc}
\frac{1}{2} & \frac{1}{2\sqrt{3}} & -\frac{1}{2} & \frac{1}{2}\sqrt{\frac{5}{3}}\\
0 & -\frac{1}{\sqrt{3}} & \frac{1}{2} & \frac{1}{2}\sqrt{\frac{5}{3}}\\
\frac{1}{2}\sqrt{3} & -\frac{1}{6} & \frac{1}{2\sqrt{3}} & -\frac{\sqrt{5}}{6}\\
0 & \frac{\sqrt{5}}{3} & \frac{1}{2}\sqrt{\frac{5}{3}} & \frac{1}{6}
\end{array}\right).
\label{eq:couplingmatrix}
\end{eqnarray}

\begin{table}
\caption{The relative partial widths of a $S$-wave and a
 $D$-wave charmonium to a pair of charmed mesons.
 The subscripts are the total spin of the two charmed meson system.}
\begin{centering}
\begin{tabular}{|c|c|c|c|c|c|c|}
\hline \hline Charmonium & $|D\bar{D}\rangle_{0}$  &
$|D^{*}\bar{D}\rangle_{1}$  & $|D\bar{D}^{*}\rangle_{1}$  &
$|D^{*}\bar{D}^{*}\rangle_{0}$  & $|D^{*}\bar{D}^{*}\rangle_{1}$  &
$|D^{*}\bar{D}^{*}\rangle_{2}$\tabularnewline \hline $S$-wave  &
$\frac{1}{12}$  & $\frac{1}{6}$  & $\frac{1}{6}$  & $\frac{1}{36}$
& 0  & $\frac{5}{9}$\tabularnewline \hline $D$-wave  &
$\frac{5}{12}$  & $\frac{5}{24}$  & $\frac{5}{24}$  & $\frac{5}{36}$
& 0  & $\frac{1}{36}$\tabularnewline \hline \hline
\end{tabular}
\par\end{centering}
\label{tab:coe}
\end{table}

As a by-product, pairs with a possible exotic quantum number can also be
obtained:
\begin{eqnarray}
|D\bar{D}^{*}+c.c.\rangle_{1^{-+}} & = & -\frac{1}{\sqrt{2}}|0\otimes1\rangle+\frac{1}{\sqrt{2}}
|1\otimes1\rangle~,\label{eq:1mp1}\\
|D^*\bar{D}^{*}\rangle^{s=1}_{1^{-+}} & = &
\frac{1}{\sqrt{2}}|0\otimes1\rangle+\frac{1}{\sqrt{2}}|1\otimes1\rangle~.\label{eq:1mp2}
\end{eqnarray}
In the heavy quark limit, the $S$-wave and $D$-wave charmonia
couple to a pair of charmed mesons through the $|1\otimes0\rangle$ and
$|1\otimes 2 \rangle$ components, respectively. Therefore, their
decay widths are proportional to the corresponding Clebsch-Gordan coefficient
squared.\footnote{In the energy region where $\psi(nS)$ dominates,
the ratio of the cross sections $e^+e^-\to D\bar{D}$, $D\bar{D}^*+c.c.$, 
$D^*\bar{D}^*_{s=0}$ and $D^*\bar{D}^*_{s=2}$ 
is also consistent with the ratio for the $S$-wave charmonium given in 
Table~\ref{tab:coe} in the HQSS limit.} 
The relative branching ratios are listed in
Table~\ref{tab:coe}.\footnote{One should notice that the ratios are
 obtained in the heavy quark limit which means the masses of $D$ and $D^*$ are 
equal to each other. In this case,
 the phase space factors of the different channels are the same and will thus 
not modify these ratios.}
This relation can also be obtained by constructing the effective
interaction based on HQSS as shown in Appendix~\ref{app:effectiveL}.

In the heavy quark limit, we can define the direct contact potentials
\begin{eqnarray}
C_{1} & \equiv & V_{01}=\langle0\otimes1|\hat{H}|0\otimes1\rangle,\quad 
C_{2}\equiv V_{10}=\langle1\otimes0|\hat{H}|1\otimes0\rangle,\label{eq:c1}\\
C_{3}&\equiv&V_{11}=\langle1\otimes1|\hat{H}|1\otimes1\rangle,\quad
C_{4}\equiv
V_{12}=\langle1\otimes2|\hat{H}|1\otimes2\rangle\label{eq:c2},
\end{eqnarray}
which are considered as constant within the small energy region
we consider. Besides the four open charm channels, i.e.
the $D\bar{D}$, $D\bar{D}^{*}+c.c.$, $D^{*}\bar{D}^{* }_{s=0}$, and
$D^{*}\bar{D}^{*}_{s=2}$ channels, there are also four expected
conventional charmonia within this region, i.e. $\psi(2S)$,
$\psi(1D)$, $\psi(3S)$, $\psi(2D)$. In the following, we use the
latin letters $i, j, \dots=1,2,3,4$ to denote the open charm
channels and the greek letters $\alpha,\beta\dots=1,2,3,4$ to denote
the bare pole terms. As a result, the corresponding potential
among these eight channels is
\begin{eqnarray}
V & = & \left(\begin{array}{cc}
v_{ij} & v_{i\beta}\\
v_{\alpha j} & 0
\end{array}\right)\label{eq:lse}
\end{eqnarray}
with
\begin{eqnarray*}
v_{ij} & = & g_{ik}^{1^{--}}g_{jk}^{1^{--}}C_{k},\quad
v_{i\beta}=g_{il}^{1^{--}}\mu_{l\beta},\quad v_{\alpha
j}=g_{jk}^{1^{--}}\mu_{ k\alpha}.
\end{eqnarray*}
Here, $\mu_{\beta l}=\mu_{l\beta}\neq0$ only when $\beta=1,3,$ and
$l=2$ or $\beta=2,4$ and $l=4$. For further use, we rename the coupling constants 
\begin{eqnarray}
\mu_{21} &\equiv& g_{2S}, \quad  \mu_{23} \equiv g_{3S},\quad
\mu_{42} \equiv g_{1D} ,\quad \mu_{44} \equiv g_{2D},
\end{eqnarray}
so that their physical meanings are manifest, i.e. these
are the couplings between the bare charmonia and the open charmed channels.

\subsection{The Lippmann-Schwinger equation}

Due to the zero component in Eq.(\ref{eq:lse}), the
Lippmann-Schwinger equation (LSE), $T=V-VGT$, can be split into two
subgroups
\begin{eqnarray}
T_{ij} &=&  V_{ij}-V_{ik}G_{k}T_{kj}-V_{i\alpha}S_{\alpha}T_{\alpha j},\label{eq:lse1}\\
T_{\alpha i}  &=&  V_{\alpha i}-V_{\alpha
j}G_{j}T_{ji},
\label{eq:lse2}
\end{eqnarray}
and
\begin{eqnarray}
T_{i\alpha} & = & V_{i\alpha}-V_{ij}G_{j}T_{j\alpha}-V_{i\beta}S_{\beta}T_{\beta\alpha}~,\label{eq:lse3}\\
T_{\alpha\beta} & = & -V_{\alpha i}G_{i}T_{i\beta}~,\label{eq:lse4}
\end{eqnarray}
with $G_{i}$ the two-body propagator and
$S_{\alpha} =(m_{\alpha}^{2}-s)^{-1}$ the bare
pole propagator. 
Substituting Eq.~(\ref{eq:lse2}) into Eq.~(\ref{eq:lse1}),
one can obtain
\begin{eqnarray}
T_{ij} & = & \hat{V}_{ij}-\hat{V}_{ik}G_{k}T_{kj}~\label{eq:les5}
\end{eqnarray}
with the effective potential $\hat{V}_{ij}=V_{ij}-V_{i\alpha}S_{\alpha}V_{\alpha
j}$.
With the above equation for $T_{ij}$ and Eq.~(\ref{eq:lse2}), one can
also calculate the transition matrix $T_{\alpha i}$ between the
charmonia and the charmed meson pair.
 Substituting Eq.~(\ref{eq:lse4}) into Eq.~(\ref{eq:lse3}), one can also extract the transition
 matrix from bare charmonia to the open charmed channels
\begin{eqnarray}
T_{i\alpha} & = &
V_{i\alpha}-\hat{V}_{ij}G_{j}T_{j\alpha}~.\label{eq:les6}
\end{eqnarray}
Again, the transition matrix $T_{\alpha\beta}$ among the charmonia 
can be extracted from Eq.~(\ref{eq:lse4})
accordingly. Now the $8\times8$ matrix is
reduced to several $4\times4$ matrices. 

The bare production amplitude is defined as
\begin{eqnarray*}
\mathcal{F} & = &
\left(F_{1,}F_{2},F_{3},F_{4},f_{1},f_{2},f_{3},f_{4}\right)^T
\end{eqnarray*}
with $f_1\equiv g_{2S}^0$, $f_2\equiv g_{1D}^0$, $f_3\equiv
g_{3S}^0$, $f_4\equiv g_{2D}^0$ the couplings between the virtual photon
and the corresponding charmonia. $F_i\equiv g^{1^{--}}_{i2}
f_S^0+g^{1^{--}}_{i4} f_D^0$ is the coupling between the virtual photon
and the $i$th open charmed channel. The physical production amplitude
can be obtained from
\begin{eqnarray*}
\mathcal{U} & = & \mathcal{F}-VG\mathcal{U}~,
\end{eqnarray*}
with the physical production amplitudes of the open charm channels and bare poles given by 
\begin{eqnarray}
\mathcal{U}_{i} & = & \mathcal{F}_{i}-V_{ij}G_{j}\mathcal{U}_{j}-V_{i\alpha}S_{\alpha}\mathcal{U}_{\alpha}~,\label{eq:pro1}\\
\mathcal{U}_\alpha&=&f_\alpha-V_{\alpha j} G_j \mathcal{U}_j.\label{eq:pro2}
\end{eqnarray}
Substituting Eq.(\ref{eq:pro2}) to Eq.(\ref{eq:pro1}), one can obtain the physical production amplitudes of the open charm channels
\bea
\mathcal{U}_i&=&\mathcal{F}_i-V_{i\alpha} S_\alpha f_\alpha -V_{ij} G_j \mathcal{U}_j+V_{i\alpha} S_\alpha V_{\alpha j} G_j \mathcal{U}_j\\
&=&\hat{\mathcal{F}}_i-\hat{V}_{ij} G_j \mathcal{U}_j
\eea
in terms of the effective bare production amplitudes $\hat{\mathcal{F}}_i=\mathcal{F}_i-V_{i\alpha} S_\alpha f_\alpha$
 and the effective potentials $\hat{V}_{ij}$. Here the contribution of the bare charmonium pole 
 is absorbed into the definition of the effective bare production amplitudes $\hat{\mathcal{F}}_i$ and the effective potentials $\hat{V}_{ij}$.

Since we only consider  separable contact potentials in our
calculation, the momentum from the two $P$-wave vertices can be
absorbed into the definition of the two-body propagator $G_i$ as
\begin{eqnarray}
G_{D\bar{D}}^{ij} & =- & 4i\int \frac{d^{4}l}{(2\pi)^{4}}
\frac{l^{i}l^{j}}{(l^{2}-m_{1}^{2}+i\epsilon)((p-l)^{2}-m_{2}^{2}+i\epsilon)}~,\label{eq:10}\\
G_{D\bar{D}^{*}}^{ij} & = & -4i\int \frac{d^{4}l}{(2\pi)^{4}}
\frac{\frac{1}{2}\varepsilon^{imn}\varepsilon^{jmk}l^{n}l^{k}}{(l^{2}-m_{1}^{2}+i\epsilon)((p-l)^{2}-m_{2}^{2}
+i\epsilon)}~,\label{eq:11}\\
G_{D^{*}\bar{D}^{*}_{s=0}}^{ij} & = & -4i\int \frac{d^{4}l}{(2\pi)^{4}}
\frac{l^{i}l^{j}}{(l^{2}-m_{1}^{2}+i\epsilon)((p-l)^{2}-m_{2}^{2}+i\epsilon)}~,\label{eq:12}\\
G_{D^{*}\bar{D}^{*}_{s=2}}^{ij} & = &
-4i\int \frac{d^{4}l}{(2\pi)^{4}}
\frac{P_{2}^{ik,mn}P_{2}^{jl,pq}\delta^{mp}\delta^{nq}l^{k}l^{l}}{(l^{2}-m_{1}^{2}+i\epsilon)((p-l)^{2}-m_{2}^{2}
+i\epsilon)}~,\label{eq:13}
\end{eqnarray}
with the $s=0$ and $s=2$ projectors
\begin{eqnarray}
P_{0}^{ij} & = & \frac{1}{\sqrt{3}}\delta^{ij},\quad
P_{2}^{ij,mn}=\sqrt{\frac{3}{5}}\left(\frac{1}{2}\delta^{im}\delta^{jn}+\frac{1}{2}\delta^{in}\delta^{jm}-\frac{1}{3}\delta^{ij}\delta^{mn}\right)
\end{eqnarray}
for the $D^*\bar{D}^*$ channel. In principle, the two additional
momentum factors from two $P$-wave vertices $l^\mu l^\nu$ can be reduced as
$\mathcal{G}_{00}g^{\mu\nu}+p^\mu p^\nu \mathcal{G}_{11}$, with $p$
the external momentum. Since the photon is produced from $e^+e^-$
annihilation, it is  transversely polarized,
$-g^{\mu\nu}+\frac{p^\mu p^\nu}{p^2}\sim \delta^{ij}$, with $i$, $j$
being spatial indexes. As a result, Eqs.~(\ref{eq:10})-(\ref{eq:13}) 
can be simplified as $-4\mathcal{G}_{00}(s,m_{i1}^2,m_{i2}^2)$, with $m_{i1}$ and $m_{i2}$ the masses
of the $i$th channels and $s$ the incoming energy squared, times the
corresponding interaction structure. The definition of
$\mathcal{G}_{00}$ can be found in Appendix~\ref{App:B00}.

\subsection{The cross section}
As shown in Fig.\ref{fig:FeynmanDiagram}, the scattering amplitude for
the process $e^+e^-\to D^{(*)}\bar{D}^{(*)}$ is
\begin{eqnarray}
\mathcal{M} & = &
\bar{v}(p_{+})(-ie\gamma_{\mu})u(p_-)\frac{iP_{\gamma}^{\mu\nu}(p)}{s}\mathcal{T}_{\nu}
\end{eqnarray}
with $p_-$ ($p_+$) the four-momentum of the electron (positron) and $p$
the sum of them. $P_{\gamma}^{\mu\nu}(p)=-g^{\mu\nu}$ is the
numerator of the photon propagator. As discussed in the above
section, since the photon has a transversal polarization, only the
transverse part $P_{\gamma}^{T\mu\nu}(p)=-g^{\mu\nu}+\frac{p^\mu
p^\nu}{p^2}$ contributes. $T^\nu$ is the production
amplitude obtained from the LSE. Then the amplitude squared is
\bea
|\mathcal{M}|^2&=&4 \frac{e^2}{s^2} \left( p_+^\nu p_-^{\nu^\prime}+p_-^\nu 
p_+^{\nu^\prime}-\frac 12 s g^{\nu\nu^\prime}\right) \mathcal{T}_\nu \mathcal{T}^*_{\nu^\prime}\\
&=&4 \frac{e^2}{s^2} \left( \frac 12 s \delta^{ij}-2 p_+^i p_+^j
\right) \mathcal{T}^i \mathcal{T}^{*j}
\eea
with $s=p^2$. The  production amplitudes for the four open charmed
channels are
\bea
\mathcal {T}^i_1&=&\mathcal{U}_1 \left( p^i_{\bar{D}}-p^i_D \right)~,\\
\mathcal{T}^i_2&=&\mathcal{U}_2\epsilon^{ijk}\left( p^j_{\bar{D}}-p^j_{D^*} \right)\varepsilon^{*k}~,\\
\mathcal {T}^i_3&=&\mathcal{U}_3 \frac{1}{\sqrt 3}\left( p^i_{\bar{D}^*}-p^i_{D^*} \right) 
\varepsilon^*_{D^*}\cdot\varepsilon^*_{\bar{D}^*}~,\\
 \mathcal{T}^i_4&=&\mathcal{U}_4 P_2^{ij,mn} \varepsilon^{*m}_{D^*} \varepsilon^{*n}_{\bar{D}^*} 
\left( p^j_{\bar{D}^*}-p^j_{D^*} \right)~.
\eea
Then the corresponding amplitudes squared are\footnote{The angular
distribution has also been discussed  in
Ref.~\cite{Voloshin:2012dk} for the bottomonium sector.}
\bea
|\mathcal{M}_{1}|^2&=&\mathcal{U}_{1}^2 \frac{32\pi\alpha}{s}|p_D|^2
(1-\cos ^2\theta)~,\label{eq:angular1}\\
|\mathcal{M}_2|^2&=&\mathcal{U}_2^2 \frac{32\pi\alpha}{s}|p_D|^2
(1+\cos ^2\theta)~,\label{eq:angular2}\\
|\mathcal{M}_{3}|^2&=&\mathcal{U}_{3}^2 \frac{32\pi\alpha}{s}|p_{D^*}|^2
(1-\cos ^2\theta)~,\label{eq:angular1}\\
|\mathcal{M}_{4}|^2&=&\mathcal{U}_{4}^2
\frac{112\pi\alpha}{5 s}|p_{D^{*}}|^2(1-\frac 17\cos
^2\theta)\label{eq:angular3}~,
\eea
with the differential cross sections given by
\bea
\frac{d\sigma_i}{d\cos\theta}=\frac{|p_{D^{(*)}}|}{64\pi
s^{3/2}}|\mathcal{M}_i|^2~.
\eea
One can obtain the total cross section by integrating over the
angle $\theta$ and then perform an  overall fit to the $e^+e^-\to
D\bar{D}$\cite{Pakhlova:2008zza}, $D\bar{D}^*$ and
$D^*\bar{D}^*$\cite{Abe:2006fj} cross sections to extract the
parameters of the model.

\begin{figure}
\begin{center}
  \includegraphics[width=0.65\textwidth]{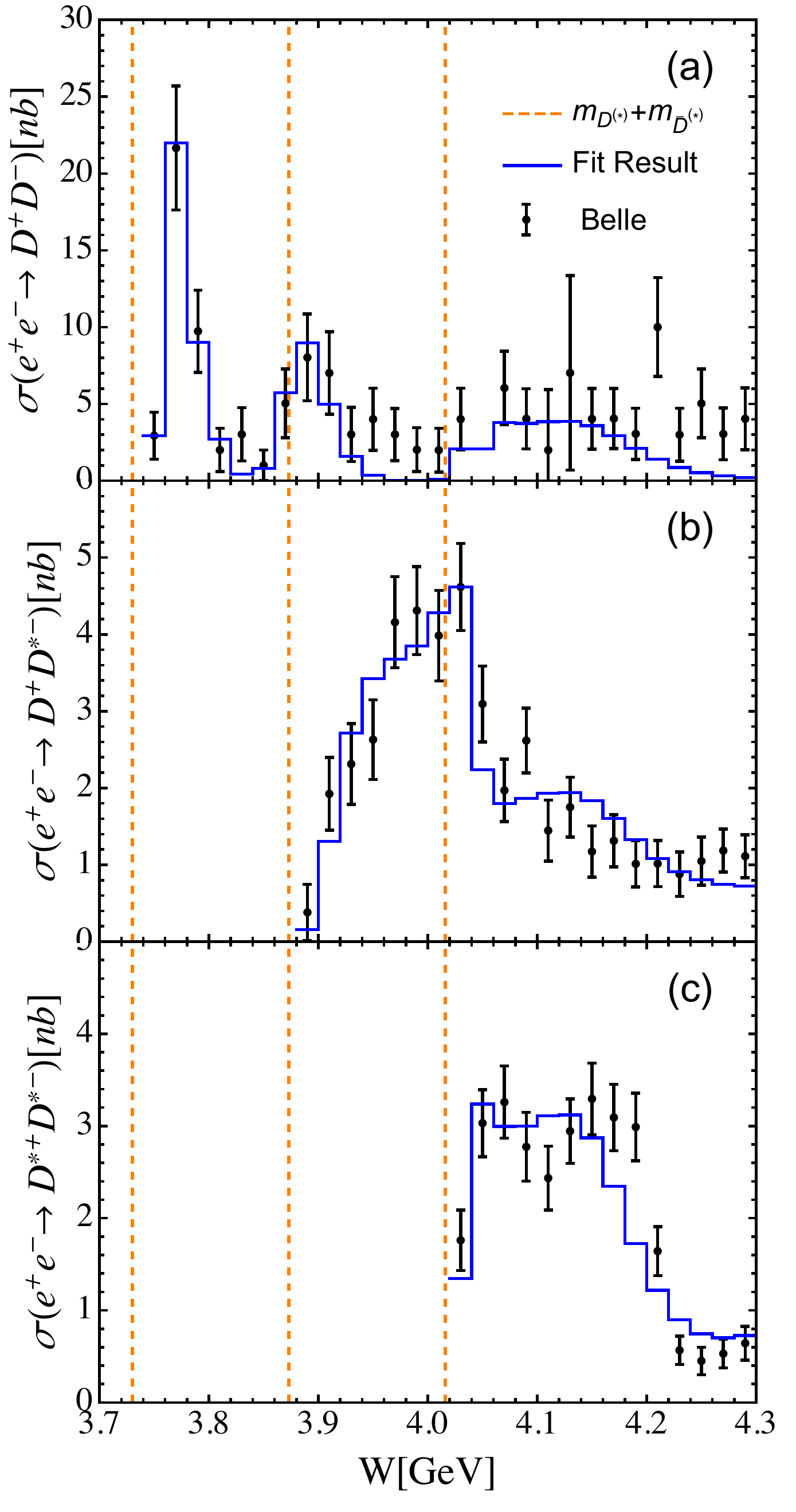}
\caption{The cross sections for $e^+e^-\to D^+D^-$, $D^+D^{*-}$, and
$D^{*+}D^{*-}$ within the energy region [3.7, 4.25]~GeV.
The three vertical lines are the $D\bar{D}$,
$D\bar{D}^*$, and $D^*\bar{D}^*$ thresholds, respectively. The
experimental data are taken from the Belle Collaboration
\cite{Pakhlova:2008zza,Abe:2006fj}.} \label{fig:CrossSection}
\end{center}
\end{figure}

\section{Results and discussions}
\label{sec:result}

In this section, we present our fit results and extract the
interesting physical quantities which can be confirmed or excluded
by  further detailed energy scans at electron-positron colliders. To obtain the 
low-energy parameters  (Table~\ref{tab:par}), an overall fit to the
$e^+e^-\to D\bar{D}$, $D\bar{D}^*$, and $D^*\bar{D}^*$ cross sections is carried out and 
and the fit results are presented in Fig.\ref{fig:CrossSection}.\footnote{Here 
and in what follows, due to the status of the experimental data, 
theoretical uncertainties are not considered but left for the forthcoming work 
after including the one-pion exchange potential and the relevant three-body
channels.} In Table~\ref{tab:par} the parameters $C_i$ of the short-ranged
contact interactions as defined in Eqs.(\ref{eq:c1}) and (\ref{eq:c2}) are displayed. 
Since  $C_2$ reflects the contribution from $|1\otimes 0\rangle$,
which has the similar effect from the $S$-wave charmonia, 
some of its contribution could be absorbed into the nearby $S$-wave charmonia.
The case for the parameter $C_4$ is analogous for the $D$-wave charmonia. 
The parameters $g_{2S}$, $g_{3S}$, $g_{1D}$, and $g_{2D}$ are
the bare couplings between the conventional $\psi(2S)$, $\psi(3S)$, 
$\psi(1D)$, $\psi(2D)$ and a pair of open charmed mesons.
$g^0_{2S}$, $g^0_{3S}$, $g^0_{1D}$, and $g^0_{2D}$ are the couplings
between the $\psi(2S)$, $\psi(3S)$, $\psi(1D)$, $\psi(2D)$ and
a virtual photon.  Since $g^0_{iS}$ and $g^0_{iD}$ are the production 
strengths of $S$-wave and $D$-wave $c\bar{c}$ through a virtual photon,
one can evaluate their ratio
\bea
\mathcal{R}_i\equiv\frac{g^0_{iD}}{g^0_{iS}}\sim \frac{(E-2 m_c)^2}{2 (E+m_c)^2}
\label{eq:r}
\eea
by plugging the plane wave Dirac spinors into the vector current $\bar{c}\gamma_\mu c$ 
with a $\mathcal{O}(\frac{\Lambda_{QCD}}{m_c})$ correction as discussed in Ref \cite{Li:2013yka}.
Here, $E$ is the total energy and $m_c$ is the charm quark mass. 
The ratio $\mathcal{R}_2=1.4\%$ is estimated with the energy at the
average of $\psi(3686)$ and $\psi(4160)$ which are expected to be
dominated by the $\psi(2S)$ and $\psi(2D)$ components, respectively. 
However, their fit values are at the same order which indicates that 
there are large corrections from higher order terms.
$f_S^0$ and $f_D^0$ are the $S$-wave and $D$-wave components 
of the couplings between a virtual photon and a pair of charmed mesons.
In principle, $f_D^0/f_S^0$ should be of the same order as
$\mathcal{R}_i$ defined in Eq.(\ref{eq:r}). However it could be
largely modified by the final-state hadronic process\cite{Wang:2013kra}, i.e.  hadronizing to
a pair of charmed mesons, which is not as good a quantity as
$\mathcal{R}_i$ to test  higher order contributions.

\begin{table}
\caption{The fit parameters.}
\vspace{2mm}
\begin{centering}
\begin{tabular}{|c|c|c|c|}
\hline\hline $C_1(\gev^{-2})$ &$C_2 (\gev^{-2}) $&$C_3 (\gev^{-2})$ & $C_4 (\gev^{-2})$ \tabularnewline \hline
$79.70\pm 1.15$ & $5.79\pm 0.22$ & $43.90\pm 0.50$ & $49.28\pm 1.37$
\tabularnewline \hline
 $g_{2S} (\gev^{0})$ & $g_{3S} (\gev^{0})$ & $ g_{1D} (\gev^{0}) $ & $ g_{2D} (\gev^{0}) $  \tabularnewline
 \hline
  $0.90\pm 0.05$ & $15.69\pm 0.04$ & $3.65\pm 0.11$ & $8.66\pm 0.15$ \tabularnewline
\hline\hline $ g^0_{2S} (\gev^{2})$ & $g^0_{3S} (\gev^{2})$ & $g^0_{1D} (\gev^{2})$ & $g^0_{2D} (\gev^{2})$
\tabularnewline \hline $0.22\pm 0.15$ & $-0.17\pm 0.01$ & $-0.05\pm
0.03$ & $-0.15\pm 0.01$ \tabularnewline \hline $f^0_S (\gev^{0})$ & $f^0_D (\gev^{0})$ &
$a (3.9~\gev)$ & $\chi^2/d.o.f.$  \tabularnewline \hline
 $-1.55\pm 0.09$ & $0.53\pm 0.08$ & $0.56\pm 0.01$ & 1.47 \tabularnewline
 \hline\hline
\end{tabular}
\par\end{centering}
\label{tab:par}
\end{table}

\subsection{The poles of {\boldmath$\psi(3770)$} and {\boldmath$\psi(4040)$}}
As shown in Fig.\ref{fig:CrossSection}, the signals of the expected
charmonia $\psi(3770)$, $\psi(4040)$ and $\psi(4160)$ are very
different in these three channels. The $\psi(3770)$ appears to 
be a pronounced peak which is isolated from other resonance structures.
In comparison, the signals for both $\psi(4040)$ and $\psi(4160)$ are not
significant in both $D\bar{D}$ and $D\bar{D}^*$ channels. They only
show some structures in the $D^*\bar{D}^*$ channel. When the
conventional charmonia are close to some thresholds, their mass
positions, line shapes or other physical quantities can be largely influenced
by their strong interactions with the open channels \cite{Eichten:2005ga}. 
In general, the $S$-wave  interactions will appear to be the most significant ones,
but sometimes the $P$-wave will also become crucial as we will show in this work.
To extract the resonance parameters of these charmonia and obtain a consistent 
understanding of the effect from the nearby thresholds, an overall fit within the coupled 
channel framework is necessary.

In the following, we extract the pole positions of the dressed charmonia on the complex 
energy plane. Since only poles which are located on the physical sheet or  close ones 
can affect the physical measurement significantly,
we only search for poles on these sheets. Accordingly, 
we only care about the poles in the following energy regions on each Riemann sheet,
\begin{align*}
\mathcal{\mathrm{I}} \quad & \mathrm{Im} ~q_{D\bar{D}}>0,\quad
\mathrm{Im}~q_{D\bar{D}^*}>0,\quad \mathrm{Im}~q_{D^*\bar{D}^*}>0,\quad\mathrm{for}~E<2 m_D~,\\
\mathcal{\mathrm{II}}\quad & \mathrm{Im} ~q_{D\bar{D}}<0,\quad\mathrm{Im}~q_{D\bar{D}^*}>0,
\quad \mathrm{Im}~q_{D^*\bar{D}^*}>0,\quad\mathrm{for}~2 m_D<E<m_D+m_{D^*}~,\\
\mathcal{\mathrm{III}} \quad & \mathrm{Im} ~q_{D\bar{D}}<0,\quad
\mathrm{Im}~q_{D\bar{D}^*}<0,\quad \mathrm{Im}~q_{D^*\bar{D}^*}>0,\quad\mathrm{for}~m_D+m_{D^*}<E<2 m_{D^*}~,\\
\mathcal{\mathrm{IV}} \quad & \mathrm{Im} ~q_{D\bar{D}}<0,\quad
\mathrm{Im}~q_{D\bar{D}^*}<0,\quad \mathrm{Im}~q_{D^*\bar{D}^*}<0,\quad\mathrm{for}~E>2 m_{D^*}~.
\end{align*}

\begin{table}
\caption{Poles on the sheets which are close to the physical sheet
and the modules of the dimensionless couplings.}
\begin{centering}
\begin{tabular}{|c|c|c|c|c|c|}
\hline\hline Sheet & Poles (GeV)&$|g_{D\bar{D}}| $ &
$|g_{D\bar{D}^*}| $ & $|g_{D^*\bar{D}^*_{s=0}}| $ &
$|g_{D^*\bar{D}^*_{s=2}}| $  \tabularnewline \hline II &
$3.764\pm i 0.006 $ & 13.53 & 9.48 & 5.88 & 16.78 \tabularnewline
\hline III & $3.879\pm i 0.035$ & 4.40 & 10.96 & 7.63 & 18.15
\tabularnewline \hline IV& $ 4.034\pm i 0.014$ & 2.90 & 2.23 & 12.52
& 12.85 \tabularnewline \hline\hline
\end{tabular}
\par\end{centering}
\label{tab:poles}
\end{table}

As shown in Table~\ref{tab:poles},\footnote{ One notices
 that the couplings in the table correspond to the relativistic fields. 
 The  non-relativistic ones can be obtained by dividing
  $\Pi_i \sqrt{2 m_i} $ with $m_i$ the masses of the heavy fields in the corresponding vertex.} the mass position of the $\psi(3770)$
on sheet~II is $11~\mev$ lower than the measured mass $3773.15\pm
0.33~\mathrm{MeV}$, which is obtained by a
Breit-Wigner fit. The fit width $12~\mathrm{MeV}$ is much smaller
than the measured width $27.2\pm 1.0~\mathrm{MeV}$. The deviation
means that there are other decay modes for the $\psi(3770)$
as discussed in Ref.\cite{Zhang:2009kr}. Another reason is that 
there is a difference between the pole width and the Breit-Wigner width 
for a broader state. The ratio of the effective
couplings to the four channels is
$13.53:9.48:5.88:16.78=1:0.70:0.43:1.24$ which is different from that of both a pure $S$-wave
charmonium $\frac{1}{2\sqrt 3}: \frac{1}{\sqrt 3}:
\frac{1}{6}: \frac{\sqrt 5}{3}=1:2:0.58:2.58$ and a pure
 $D$-wave charmonium $\frac 12\sqrt{\frac 53}: \frac 12\sqrt{\frac
53}:\frac{\sqrt 5}{6}:\frac 16=1:1:0.58:0.26$. One might expect that 
the ratios should lie within the range limited by the values of the pure $S$-wave charmonium
 and pure $D$-wave charmonium. However, for instance, $0.70$ is smaller than both $2$ ($S$-wave case) and $1$ ($D$-wave case).
  When the masses of $D$ and $D^*$ are set equal to each other, 
the ratios of the couplings lie between those of the pure $S$-wave charmonium 
and the pure $D$-wave charmonium as one expects. It indicates that the deviation is due to
the HQSS breaking effects stemming from the mass splitting between $D$ and $D^*$ .
Since it strongly couples to $D^*\bar{D}^*_{s=2}$ channel with the $|1\otimes 0\rangle$ as 
the dominant component, the most important hidden charm decay channel is $J/\psi$ plus two $S$-wave 
pions.\footnote{PDG gives $\mathcal{BR}(\psi(3770)\to J/\psi \pi^+\pi^-)=(1.93\pm 0.28)\times 10^{-3}$ and
$\mathcal{BR}(\psi(3770)\to J/\psi \pi^0\pi^0)=(8.0\pm 3.0)\times 10^{-4}$ which is larger than 
that of the other hidden charm decay channels.}

We also find a pole $4.032\pm i 0.016$~GeV on sheet~IV which corresponds to the $\psi(4040)$ with 
the ratio of the effective couplings $2.90:2.23:12.52:12.85=1:0.77:4.32:4.43$.
It does not agree with either the ratio given by the pure $S$-wave or pure $D$-wave 
charmonia which also indicates large HQSS breaking effects.
Since it couples to $D^*\bar{D}^*_{s=0}$ and $D^*\bar{D}^*_{s=2}$ channels which are 
dominated by the $|0\otimes 1\rangle$ and $|1\otimes 0\rangle$ components with relatively large 
strengths, respectively, its favored hidden charm decay channels would be $\eta_c$ plus a
 light isosinglet vector (such as $\omega$ or three pions) and $J/\psi$ plus two $S$-wave pions.
 
 The poles corresponding to $\psi(3686)$ and $\psi(4160)$ are not listed in Table~\ref{tab:poles}.
  As the energy of $\psi(3686)$ is below the open charm thresholds,
   it mainly decays into light mesons and lower charmonia,
   in which case our method lacks precision. On the other side, 
   $\psi(4160)$ is much more influenced by higher thresholds
   which we will investigate in detail in a forthcoming work.

\subsection{The interpretation of the {\boldmath$G(3900)$}}
Since the observation of the $G(3900)$ in the ISR process $e^+e^-\to
(\gamma) D\bar{D}$ by BABAR and Belle
\cite{Aubert:2006mi,Pakhlova:2008zza} in 2007 and 2008,
 various groups have been paying attention to the $P$-wave
interaction between a pair of charmed mesons   \cite{Zhang:2010zv,Chen:2012qq,Liu:2010xh,Achasov:2012ss,Cao:2014qna,Eichten:1979ms}. 
The $P$-wave interaction between $D$ and $D^*$ is of great
interest since it can be investigated in $e^+e^-$ collisions.
With the fit parameters, we find a pole at $3.879\pm i 0.032~\gev$ on sheet~III
$4~\mev$ above the $D\bar{D}^*$ threshold, which corresponds to the $G(3900)$ structure.
It is a resonance
and can decay into the two lower channels, i.e. $D\bar{D}$ and
$D\bar{D}^*$. Its coupling ratio to the four channels is
$4.40:10.96:7.63:18.15=1:2.49:1.73:4.13$. Although it deviates from
the ratios by the pure $S$-wave  and pure $D$-wave
charmonium, we cannot conclude that it does not have any $c\bar{c}$
component due to the HQSS breaking.
 Because it strongly couples to the $D\bar{D}^*$ and $D^*\bar{D}^*_{s=2}$ channels,
it will show a significant threshold effect at the $D\bar{D}^*$ and
$D^*\bar{D}^*$ thresholds, especially in the isospin breaking
channels\cite{Wang:2011yh} due to the mass difference between 
the charged meson loops and the neutral ones. Since the $|1\otimes 2\rangle$ and
$|1\otimes 0\rangle$ components dominate the $D\bar{D}^*$ and
$D^*\bar{D}^*_{s=2}$ channels, respectively, we would also expect
its signal in $J/\psi$ plus two $D$-wave pions and $S$-wave pions
channels. To further pin down the nature of the $G(3900)$, 
higher-statistics data are necessary.

\subsection{The angular distribution in {\boldmath$D^*\bar{D}^*$} channels}
As discussed in Sec.~\ref{sec:decomposition}, the sum of the $D^*\bar{D}^*$
spins can either be zero or two for the $J^{PC}=1^{--}$ channel. This corresponds 
to the angular distribution $1-\cos^2\theta$, Eq.(\ref{eq:angular1}), and 
$1-\frac17\cos^2\theta$ [Eq.~(\ref{eq:angular3})], respectively. Since these
two bases, i.e. the $D^*\bar{D}^*_{s=0}$ and $D^*\bar{D}^*_{s=2}$
bases, are orthogonal to each other, the events in the $D^*\bar{D}^*$
channel are the incoherent sum of these. The different angular
distribution means that it can help  to disentangle how large the
$s=0$ and $s=2$ components should be and can also be viewed as evidence
for our scenario.

Since the $1-\frac 17\cos^2\theta$ distribution cannot be
distinguished from a flat distribution if the integrated luminosity is not high enough, one can alternatively
define an asymmetry parameter
\bea
\mathcal{A}(E)\equiv\frac{\int_{-1.0}^{-0.5}
\frac{d\sigma(E)}{d\cos\theta} ~d\cos\theta+\int_{0.5}^{1.0}
\frac{d\sigma(E)}{d\cos\theta} ~d\cos\theta}{\sigma(E)}
\eea
to disentangle the components. As discussed above, since the
$\psi(4160)$ will be affected by large effects from the next threshold, the
angular distribution at $4.04~\mathrm{GeV}$ is the best energy point to
disentangle it. Our results indicate that the ratio between these two components 
at $4.04~\mathrm{GeV}$ is
\bea
\frac{d\sigma_{s=0}/d\cos\theta}{d\sigma_{s=2}/d\cos\theta} =
\frac{0.41(1-\cos^2\theta)}{0.23(1-\frac 17 \cos^2\theta)},
\eea
with the total cross sections $1.52~\nb$ for
$s=0$ and $1.23~\nb$ for $s=2$.
 With our fit parameters, the asymmetry is estimated as
\bea
\mathcal{A}(4.04)=0.39.
\eea
For pure $s=0$ and  pure $s=2$, the asymmetries are $ \mathcal{A}_{s=0}(4.04)=0.31$ and $ \mathcal{A}_{s=2}(4.04)=0.48$, respectively.
It can be confirmed or excluded by a  further detailed energy scan at
BESIII with high integrated luminosity.

\subsection{Searching the {\boldmath$1^{-+}$} exotic candidate}
Besides the $1^{--}$ quantum number for the $P$-wave interaction,
 $1^{-+}$ quantum number can also be obtained~\cite{Wang:2014wga} as shown in 
Eqs.~(\ref{eq:1mp1}) and (\ref{eq:1mp2}). Since $1^{-+}$ is an exotic quantum number which 
cannot be obtained with a conventional $c\bar{c}$ configuration,
there is no bare charmonium pole contribution in this channel. The only relevant low-energy parameters are
$C_1$ and $C_3$ due to the appearance of the $|0\otimes 1\rangle$
and $|1\otimes 1\rangle$ components in the wave functions of the $1^{-+}$ channel.
As a result, the corresponding contact potential is
 \bea
 V^{1^{-+}}=\left ( \begin{array}{cc}
 \frac 12 C_1+\frac 12 C_3 &  -\frac 12 C_1+\frac 12 C_3\\
  -\frac 12 C_1+\frac 12 C_3 &  \frac 12 C_1+\frac 12 C_3
 \end{array}\right)~,
 \label{eq:v1mp}
 \eea
 with the values of $C_1$ and $C_3$ given in Table~\ref{tab:par}. Since the relevant 
channels are $D\bar{D}^*$  and $D^*\bar{D}^*$, we can define the physical Riemann sheets and 
the close sheets as
 \begin{align*}
\mathcal{\mathrm{I}} \quad & \quad\mathrm{Im}~q_{D\bar{D}^*}>0,\quad 
\mathrm{Im}~q_{D^*\bar{D}^*}>0,\quad \mathrm{for} ~  E<m_D+m_{D^*}~,\\
\mathcal{\mathrm{II}}\quad &\quad\mathrm{Im}~q_{D\bar{D}^*}<0,\quad \mathrm{Im}~q_{D^*\bar{D}^*}>0,\quad
 \mathrm{for} ~ m_D+m_{D^*}<E<2 m_{D^*}~,\\
\mathcal{\mathrm{III}} \quad & \quad\mathrm{Im}~q_{D\bar{D}^*}<0,\quad 
\mathrm{Im}~q_{D^*\bar{D}^*}<0, \quad \mathrm{for} ~ E>2 m_{D^*}~.
\end{align*}
With the central values of  $C_1$ and $C_3$, we obtain two poles which would affect the measurements. 
One is on the physical sheet  about $500~\mev$
below the $D\bar{D}^*$ threshold and couples to the
$D\bar{D}^*$ and $D^*\bar{D}^*_{s=1}$ channels with equal strength. 
 The equal couplings are a consequence of the two facts of the HQSS. 
One is that the two diagonal elements in the potential
[Eq.~(\ref{eq:v1mp})] are the same and similarly for those two off-diagonal elements.
 Another one is that the mass splitting between $D$ and $D^*$ is much smaller 
 than the difference between the pole mass and the open-charm thresholds. 
 In this case, the masses of $D$ and $D^*$ are approximately degenerate. 
 No matter on which sheets the poles are located, once they are
below the two thresholds (bound states on the physical sheet or virtual states on an 
unphysical sheet), they  couple with equal strength to these two channels. 
 However, since the pole is very deep with about
$500~\mev$ binding energy, it is far beyond the applicable energy
region of the contact term interaction. One cannot predict reliably any deep pole without
energy-dependent potentials, such as the one-pion exchange potential.

\begin{table}
\caption{Poles for the $1^{-+}$ channels on the physical sheet and
the close sheets as well as their dimensionless couplings to each channel. }
\vspace{2mm}
\begin{centering}
\begin{tabular}{|c|c|c|c|}
\hline\hline Sheets &Poles (GeV)& $|g_{D\bar{D}^*}|$ &
$|g_{D^*\bar{D}^*_{s=1}}|$  \tabularnewline 
 \hline II & $3.915\pm i 0.003$ & 7.91
& 3.48  \tabularnewline \hline\hline
\end{tabular}
\par\end{centering}
\label{tab:poles2}
\end{table}

We also find a pole $3.915\pm i 0.003~\gev$ on sheet~II with the
effective couplings $7.91$ and $3.48$ to the $D\bar{D}^*$ and
$D^*\bar{D}^*$ channels, respectively. It is only $40~\mev$ above
the $D\bar{D}^*$ threshold which is acceptable with only the
short-ranged contact interaction.  The $|0\otimes 1\rangle$
component has two possible $S$-wave decay channels dictated by the 
symmetry. The product can be either $h_c+(3\pi)_{1^{--}}$ with the isosinglet  $J^{PC}=1^{--}$ three pions\footnote{Here 
and in what follows, the quantum number $J^{PC}$ of pions are explicitly written as subindices.}
or $\eta_c+(4\pi)_{1^{++}}$. However, only the $\eta_c+(4\pi)_{1^{++}}$ 
can be accessed due to the phase space constraint. 
The $|1\otimes 1\rangle$ component can
$P$-wave decay to both $J/\psi+(3\pi)_{1^{--}}$ 
and $\chi_{cJ}+(2\pi)_{0^{++}}$. Among those $J/\psi$ plus three pions 
channel is the most favoured one due to its larger phase space. Therefore, we would 
expect that the $J/\psi +(3\pi)_{1^{--}}$ and $\eta_c +(4\pi)_{1^{++}}$ channels are the best ones to measure this 
potential $1^{-+}$ exotic state in the $e^+e^-\to\gamma J/\psi 3\pi$ and $e^+e^-\to \gamma \eta_c 4\pi$ processes.

The wave functions of the two $1^{-+}$ exotic states, i.e.  Eqs.(\ref{eq:1mp1}) and (\ref{eq:1mp2}) ,
are similar to those of the $Z_b(10610)$ and $Z_b(10650)$, cf. Eq.~(3) in Ref.~\cite{Bondar:2011ev}.
The similarity means that the signs of the component $|1\times 1 \rangle$ ($|0 \times 1\rangle$) 
with heavy quark spin $s_Q=1$ ($s_Q=0$) in these two wave functions 
are the same (opposite). Consequently, the production amplitudes of these two $1^{-+}$ states in
$e^+e^-\to\gamma 1^{-+}\to\gamma J/\psi 3\pi$ and $e^+e^-\to\gamma 1^{-+}\to\gamma \eta_c 4\pi$ processes 
are
  \bea
\mathcal{A}(e^+e^-\to\gamma 1^{-+}\to\gamma J/\psi 3\pi)&\propto&  
\frac{g_{\gamma 1} g_{J/\psi 1}}{E-E_1+i\Gamma_1/2}+\frac{g_{\gamma 2} g_{J/\psi 2}}{E-E_2+i\Gamma_2/2}~, 
\label{eq:a1}\\
 \mathcal{A}(e^+e^-\to\gamma 1^{-+}\to\gamma \eta_c 4\pi)&\propto&  
\frac{g_{\gamma 1} g_{\eta_c 1}}{E-E_1+i\Gamma_1/2}+\frac{g_{\gamma 2} g_{\eta_c 2}}{E-E_2+i\Gamma_2/2}~, \label{eq:a2}
 \eea
with $g_{\gamma i}$ the coupling strengths between the $i$th $1^{-+}$ state and two photons. 
$E_{i}$ and $\Gamma_i$ are the energy and width of the $i$th $1^{-+}$ state, respectively. 
In the HQSS limit, $g_{J/\psi 2}/g_{J/\psi 1}=g_{\gamma 2}/g_{\gamma 1}=1$ due to the same sign of 
the component $|1\times 1 \rangle$ in Eqs.~(\ref{eq:1mp1}) and (\ref{eq:1mp2}). 
The case for the $|0\times 1 \rangle$ component is different, i.e.  $g_{\eta_c 2}/g_{\eta_c 1}=-1$.
One would expect that there is  destructive (constructive) interference 
in the $J/\psi 3\pi$ ($\eta_c 4\pi$) invariant mass distribution  when the energy lies 
between the pole positions, which is similar to that for the two $Z_b$ states
in the $\Upsilon\pi$ ($h_b\pi$) channel. However, the two poles in our case are not close enough to make
this interference pattern as significant as that for the two $Z_b$ states \cite{Bondar:2011ev}. 

\section{Summary}\label{sec:summary}
We have studied the $P$-wave $D\bar{D}$, $D\bar{D}^*$, and $D^*\bar{D}^*$
coupled channel effects with a the short-ranged separable contact term interaction
in the heavy quark limit by solving the Lippmann-Schwinger equation.
To extract the physical quantities, we fit the cross sections of $e^+e^-\to D\bar{D}$,
$D\bar{D}^*$, and $D^*\bar{D}^*$ within the energy region $[3.7,
4.25]~\gev$. Since there are some conventional charmonia, such as
$\psi(2S)$, $\psi(3S)$, $\psi(1D)$, and $\psi(2D)$ in this mass region, 
these bare pole terms are also included in 
the calculation in addition to the contact potential. After having fitted the parameters
of the model, we extract the pole positions
of the $\psi(3770)$ and $\psi(4040)$ as well as their couplings to each
channel. The pole positions of both $\psi(3686)$ and $\psi(4160)$ 
are left for the forthcoming work after including all the relevant thresholds.
It is an efficient and consistent way to extract the resonance parameters and the
couplings of a state below threshold. Besides the two poles, another pole at
$3.879\pm i 0.035~\gev$ on the unphysical sheet is also found which may
correspond to the so-called $G(3900)$ observed by
BABAR and Belle. However, due to the limitations in statistics of these
data, to further pin down the nature of $G(3900)$ a detailed scan of the open
charmed channels in the $e^+e^-$ annihilation process is necessary. 
We also propose that the angular distribution, or the asymmetry
$\mathcal{A}$ if the luminosity is not high enough, in
the $D^*\bar{D}^*$ channel can test our model.

Besides the $1^{--}$ quantum number, the $P$-wave $D\bar{D}^*$ and
$D^*\bar{D}^*$ interaction can also access the exotic $1^{-+}$
quantum number. Since $1^{-+}$ is an exotic quantum number, there
is no bare pole term in this channel. With the relevant parameters
$C_1$ and $C_3$ fitted in the $1^{--}$ channels, we find a pole
$3.915\pm i 0.003~\gev$, $40\mev$ above $D\bar{D}^*$ threshold. It can
be measured in the $J/\psi +3\pi$ and $\eta_c +4\pi$ invariant mass
distributions in $e^+e^-\to \gamma J/\psi 3\pi$ and $e^+e^-\to
\gamma \eta_c 4 \pi$ processes by further experiments with high
integrated luminosity.

\medskip

\section*{ACKNOWLEDGMENTS}
We give thanks to Feng-Kun Guo, Zhi-Hui Guo, Christoph Hanhart,
and  Qiang Zhao for useful discussions and comments. This work is
supported in part by the DFG and the NSFC through funds provided to
the Sino-German CRC 110 ``Symmetries and the Emergence of Structure
in QCD.'' The work of U. G. M. was also supported by the Chinese Academy 
of Sciences (CAS) President's International Fellowship Initiative (PIFI) 
(Grant No. 2015VMA076).


\medskip

\begin{appendix}

\section{THE EFFECTIVE LAGRANGIAN OF $S$-WAVE AND $D$-WAVE CHARMONIA COUPLING TO A PAIR OF CHARMED MESONS}
\label{app:effectiveL} Due to the heavy quark spin symmetry, the
$S$-wave charmed mesons, i.e. $D$ and $D^*$, can be collected as a
spin doublet superfield $H_{a}=V_{a}^{i}\sigma^{i}+P_{a}$ which
annihilates the corresponding $Q\bar{q}$ charmed mesons. Here, $V$
and $P$ denote vector and pseudoscalar charmed mesons, respectively. 
Its charged conjugate partner is
$\bar{H}_{a}=\sigma_{2}\mathcal{C}H_{a}^{T}\mathcal{C}^{-1}\sigma_{2}=-\bar{V}_{a}^{i}\sigma^{i}+\bar{P}_{a}$
with the convention $\mathcal{C}V\mathcal{C}^{-1}=\bar{V}$ and
$\mathcal{C}P\mathcal{C}^{-1}=\bar{P}$. The corresponding creation
superfield is
\begin{equation}
H_{a}^{\dagger}=V_{a}^{i\dagger}\sigma^{i}+P_{a}^{\dagger},\quad\bar{H}_{a}^{\dagger}
= -\bar{V}_{a}^{i\dagger}\sigma^{i}+\bar{P}_{a}^{\dagger}.
\end{equation}
In the heavy quark limit, the $S$-wave and $D$-wave charmonia can
also be collected in multiplets $J=\psi^i\sigma^i+\dots$ and
\begin{eqnarray}
J^{ij}&=&\frac 12 \sqrt{\frac 35} (\sigma^i
\psi_{D1}^j+\sigma^j\psi_{D1}^i)-\frac {1}{\sqrt{15}}
\delta^{ij}\vec{\sigma}\cdot \vec{\psi}_{D1}+\ldots~.
\label{eq:LagDwave1}
\end{eqnarray}
Since we consider the $1^{--}$ channel, only the relevant vector
charmonia are presented explicitly. The Lagrangian of the $S$-wave
and $D$-wave charmonia coupling to a pair of charmed mesons reads
\begin{eqnarray}
\mathcal{L} & = & i\frac{g_{S}}{2}\langle
J^{\dag}H_{a}\sigma^{i}\partiallr^{i}\bar{H}_{a}\rangle+i\frac{g_{D}}{2}\langle
J^{ij\dag}H_{a}\sigma^{i}\partiallr^{j}\bar{H}_{a}\rangle+\text{H.c.}
\end{eqnarray}
with the overall coupling strength $g_S$ and $g_D$. Using the above
interaction, one can also get the same branching ratio fractions as
shown in Table~\ref{tab:coe}.

\section{LOOP FUNCTIONS}\label{App:B00}
In this appendix, we present the relevant one-loop integrals explicitly.
We use the standard tensor reduction \cite{Passarino:1978jh} to express the occurring
integrals as a linear sum of scalar one-loop functions.
Here we only present the one-loop function we need in the calculation \cite{Yao:2015qia}.
The other two-point one-loop functions can be found in Ref.\cite{Yao:2015qia}.

We use
\bea
R=\frac{2}{d-4}+\gamma_E-\ln(4\pi)\ ,
\eea
to denote the ultraviolet divergences
with $\gamma_E$  the Euler constant and $d$ the number of space-time dimensions. 
The one-point function is defined as
 \bea
 \mathcal{I}_0(M_a^2)=\frac{\mu^{4-d}}{i}\int\frac{\textrm{d}^dk}{(2\pi)^d}
\frac{1}{k^2-M_a^2+i0^+}=-\frac{M_a^2}{16\pi^2}
\left(R-1+\ln\frac{M_a^2}{\mu^2}\right) .
 \eea
 The second rank tensor loop can be decomposed as
 \bea\nonumber
 \mathcal{G}^{\mu\nu}(p^2,M_a^2,M_b^2)&=&\frac{\mu^{4-d}}{i}\int\frac{\textrm{d}^dk}{(2\pi)^d} \frac{k^{\mu}
k^{\nu} }{\left(k^2-M_a^2+i0^+\right)\left[(k+p)^2-M_b^2+i0^+\right]}\\
&=&g^{\mu\nu} \mathcal{G}_{00}(p^2,M_a^2,M_b^2)+p^\mu p^\nu \mathcal{G}_{11}(p^2,M_a^2,M_b^2)~,
 \eea
 where
 \bea\nonumber
 \mathcal{G}_{00}(p^2,M_a^2,M_b^2)&=&\frac{1}{12p^2}((p^2+\Delta_{ab})
\mathcal{I}_0(M_a^2)+(p^2-\Delta_{ab})\mathcal{I}_0(M_b^2) \\\nonumber
 &+& \left[4p^2\,M_a^2-(p^2+\Delta_{ab})^2\right] \mathcal{G}(p^2,M_a^2,M_b^2) )
-\frac{1}{16\pi^2}\frac{1}{18}(p^2-3\Sigma_{ab})~,\
 \eea
with $\Delta_{ab}
\equiv M_a^2-M_b^2$ and $\Sigma_{ab} \equiv M_a^2+M_b^2$. 
 In the heavy quark symmetry limit, i.e. 
$m_D = m_{D^*}$, the $\mathcal{I}_0(M^2)$ part would be a constant which could be absorbed 
into the contact terms $C_i$. Further, the fundamental loop integral
$\mathcal{G}(s,M_a^2,M_b^2)$ reads
\bea
\mathcal{G}(s,M_a^2,M_b^2) \al=\al
\frac{1}{16\pi^2}\big\{ a(\mu)-\ln\frac{M_a^2}{\mu^2}-\frac{s-M_a^2+M_b^2}{2s}\ln\frac{M_b^2}{M_a^2} \\
 \al \al -\frac{\sigma_{ab}}{2s}\big[  \ln (s-M_b^2+M_a^2+\sigma_{ab})-\ln (-s+M_b^2-M_a^2+\sigma_{ab}) \\
\al \al  +\ln (s+M_b^2-M_a^2+\sigma_{ab}) -\ln (-s-M_b^2+M_a^2+\sigma_{ab})\big] \big\}
\eea
with $\sigma_{ab} = \sqrt{[s-(M_a+M_b)^2][s-(M_a-M_b)^2]}$
and $a(\mu)$ the substraction constant which depends on the  scale of dimensional regularization $\mu$.
\end{appendix}


\end{document}